\documentclass{emulateapj}
\usepackage{graphicx}

\shortauthors{Fixsen et al.}
\shorttitle{CMB Temperature at 10 GHz}
\slugcomment{Submitted to The Astrophysical Journal}

\begin{document}

% Needed TeX macros
\def\COBE{{\sl COBE\/}}
\def\wisk#1{\ifmmode{#1}\else{$#1$}\fi}
\def\um     {\wisk{{\rm \mu m}}}
\def\etal   {et~al.}
\def\deg    {\wisk{^\circ\ }}
\def\temp   {2.721~K}
\def\err    {$\pm~10$~mK}

% ---------------------Title ---------------------
\title{The Temperature of the CMB at 10 GHz}

% --------------------- Author list ---------------------
\author{ D.J. Fixsen\altaffilmark{1}, A. Kogut\altaffilmark{2},
S. Levin\altaffilmark{3}, M. Limon\altaffilmark{1}, 
P. Lubin\altaffilmark{4},P. Mirel\altaffilmark{1},
M. Seiffert\altaffilmark{3}, and E. Wollack\altaffilmark{2} }

\altaffiltext{1}{SSAI, 
                 Code 685, NASA/GSFC, 
                 Greenbelt MD 20771.
		 e-mail: fixsen@stars.gsfc.nasa.gov}
\altaffiltext{2}{Laboratory for Astronomy and Solar Physics,
                 Code 685, NASA/GSFC, 
                 Greenbelt MD 20771.}
\altaffiltext{3}{Jet Propulsion Laboratory, California Institute of Technology
		4800 Oak Grove Drive, 
		Pasadena, CA 91109.}
\altaffiltext{4}{University of California at Santa Barbara}

% --------------------- Abstract ---------------------
\begin{abstract}
We report the results of an effort to measure the low frequency portion of 
the spectrum of the Cosmic Microwave Background Radiation (CMB), using a 
balloon-borne instrument called ARCADE (Absolute Radiometer for Cosmology, 
Astrophysics, and Diffuse Emission). These measurements are to search for 
deviations from a thermal spectrum that are expected to exist in the CMB due 
to various processes in the early universe.

The radiometric temperature was measured at 10 and 30~GHz using a 
cryogenic open-aperture instrument with no emissive windows. An external
blackbody calibrator provides an {\it in situ} reference. A linear model is 
used to compare the radiometer output to a set of thermometers on the 
instrument. The unmodeled residuals are less than 50~mK peak-to-peak with a 
weighted RMS of 6~mK. Small corrections are made for the residual emission 
from the flight train, atmosphere, and foreground Galactic emission. 
The measured radiometric temperature of the CMB is 2.721$\pm .010$~K at 10~GHz 
and 2.694$\pm 0.032$~K at 30~GHz.
\end{abstract}


% --------------------- Main text ---------------------

\section{INTRODUCTION} 
Since the discovery of the CMB a key question has been: How does it deviate
from a perfect uniform black body spectrum? The FIRAS (Far InfraRed Absolute
Spectrophotometer) instrument (Fixsen \& Mather 2002, Brodd \etal\ 1997) has 
shown the spectrum is nearly an ideal black body spectrum from $\sim60$~GHz 
to $\sim600$~GHz with temperature 2.725$\pm$.001 K. At lower frequencies the 
spectrum is not so tightly constrained; plausible physical processes 
(reionization, particle decay) could generate detectable distortions
below 10 GHz while remaining undetectable by the FIRAS instrument. The ARCADE 
(Absolute Radiometer for Cosmology, Astrophysics and Diffuse Emission) 
experiment observes the CMB spectrum at frequencies a decade below FIRAS
to search for potential distortions from a blackbody spectrum.

The frequency spectrum of the cosmic microwave background (CMB) carries a 
history of energy transfer between the evolving matter and radiation fields
in the early universe. Energetic events in the early universe
(particle decay, star formation) heat the diffuse matter which then cools 
via interactions with the background radiation, distorting the radiation 
spectrum away from a blackbody. The amplitude and shape of the resulting 
distortion depend on the magnitude and redshift of the energy transfer
(Burigana \etal\ 1991, Burigana \etal\ 1995).

The primary cooling mechanism is Compton scattering of hot electrons against 
a colder background of CMB photons, characterized by the dimensionless integral
\begin{equation}
y = \int^z_0 ~\frac{ k[T_e(z) - T_{\gamma}(z)] }{ m_e c^2} 
\sigma_T n_e(z) c \frac{dt}{dz^\prime} dz^\prime,
\label{compton_y_definition}
\end{equation}
of the electron pressure $n_e k T_e$ along the line of sight, where $m_e$, 
$n_e$ and $T_e$ are the electron mass, spatial density, and temperature,
$T_{\gamma}$ is the photon temperature, $k$ is Boltzmann's constant, $z$ is redshift,
and $\sigma_T$ denotes the Thomson cross section (Sunyaev \& Zeldovich 1970).
For recent energy releases $z < 10^4$, the gas is optically thin, resulting 
in a uniform decrement $ \Delta T_{\rm RJ} = T_{\gamma} (1 - 2y) $
in the Rayleigh-Jeans part of the spectrum where there are too few photons, 
and an exponential rise in temperature in the Wien region with too many 
photons. The magnitude of the distortion is related to the total energy transfer
\begin{equation}
\frac{\Delta {\rm E}}{\rm E} = {\rm e}^{4y} - 1 \approx 4y
\label{delta_e_vs_y_eq}
\end{equation}
Energy transfer at higher redshift $10^4 < z < 10^7$ approaches the 
equilibrium Bose-Einstein distribution, characterized by the dimensionless 
chemical potential $ \mu_0 = 1.4 \frac{\Delta {\rm E}}{\rm E}. $
Free-free emission thermalizes the spectrum at long 
wavelengths. Including this effect, the chemical potential becomes frequency-dependent,
\begin{equation}
\mu(x) = \mu_0 \exp(- \frac{2x_b}{x}),
\label{mu_vs_freq_eq}
\end{equation}
where $x_b$ is the transition frequency at which Compton scattering of photons 
to higher frequencies is balanced by free-free creation of new photons.
The resulting spectrum has a sharp drop in brightness temperature at centimeter 
wavelengths (Burigana \etal\ 1991). A chemical potential distortion would 
arise, for instance, from the late decay of heavy particles
produced at much higher redshifts.

Free-free emission can also be an important cooling mechanism.
The distortion to the present-day CMB spectrum is given by
\begin{equation}
\Delta T_{\rm ff} = T_{\gamma} \frac{Y_{\rm ff}}{x^2}
\label{FF_distortion_eq}
\end{equation}
where $x$ is the dimensionless frequency $h \nu / k T_{\gamma}$,
$Y_{\rm ff}$ is the optical depth to free-free emission
\begin{equation}
Y_{\rm ff} = \int^z_0 ~\frac{ k[T_e(z) - T_{\gamma}(z)] }{ T_e(z) }
\frac{ 8 \pi e^6 h^2 n_e^2 g }{ 3 m_e (kT_{\gamma})^3 \sqrt{6\pi m_e k T_e} }
\frac{dt}{dz^\prime} dz^\prime,
\label{Yff_definition}
\end{equation}
and g is the Gaunt factor (Bartlett \& Stebbins 1991). The distorted CMB 
spectrum is characterized by a quadratic rise in temperature at long 
wavelengths. Such a distortion is an expected signal of the reionization 
of the intergalactic medium by the first collapsed structures.

%-----------------------------------------------------------
% Figure 1: Current limits to CMB distortions
%-----------------------------------------------------------
\begin{figure}
\includegraphics[angle=90,width=3.25in]{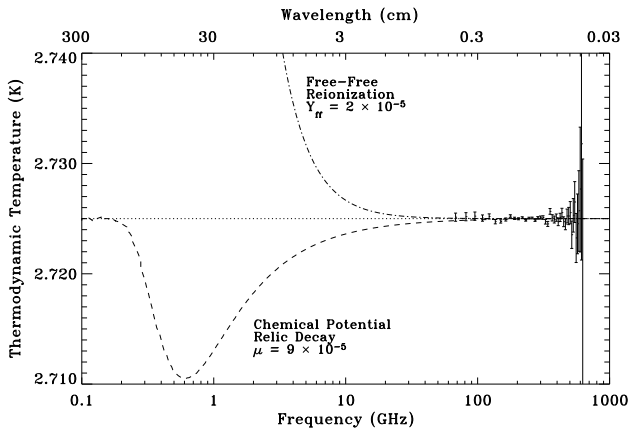}
\caption{Current 95\% confidence upper limits to distorted CMB spectra.
Measurements at short wavelengths (Fixsen \etal\ 1996)
do not preclude detectable signals at wavelengths longer than 1 cm.
\label{arcade_vs_firas} }
\end{figure}

Figure \ref{arcade_vs_firas} shows current upper limits to CMB spectral 
distortions. Measurements at wavelengths shorter than 1~cm are consistent 
with a blackbody spectrum, limiting $y < 14 \times 10^{-6}$ and
$\mu < 9 \times 10^{-5}$ at 95\% confidence (Fixsen \etal\ 1996, 
Gush \etal\ 1990). Direct observational limits at longer wavelengths are weak.
Reionization is expected to produce a cosmological free-free background
with amplitude of a few mK at frequency 3 GHz (Haiman \& Loeb 1997, Oh 1999).
The most precise observations (Table \ref{measure})
have uncertainties much larger than the predicted signal from reionization.
Existing data only constrain $|Y_{\rm ff}| ~< ~1.9 \times 10^{-5}$,
corresponding to temperature distortions $\Delta T < 19 $ mK at 3 GHz
(Bersanelli \etal\ 1994).

Uncertainties in previous measurements have been dominated by systematic 
uncertainties in the correction for emission from the atmosphere, Galactic 
foregrounds, or warm parts of the instrument. ARCADE represents a long-term 
effort to improve measurements at cm wavelengths using a cryogenic 
balloon-borne instrument designed to minimize these systematic errors.
This paper presents the first results from the ARCADE program.

\section{THE INSTRUMENT}
The ARCADE is a balloon-borne instrument with two radiometers at 10 and 30 GHz 
mounted in a liquid helium dewar. Each radiometer consists of cryogenic and 
room temperature components. A corrugated horn antenna, a Dicke switch 
consisting of a waveguide ferrite latching switch, an internal reference load 
constructed from a waveguide termination, and GaAs HEMT amplifier comprise the 
cryogenic components. The signal then passes to a 270 K section consisting of 
additional amplification and separation into two sub-bands followed by
diode detectors, making four channels in all (Kogut \etal\ 2004a). In addition 
there is an external calibrator which can be positioned to fully cover 
the aperture of either the 10 or 30~GHz horn (Kogut \etal\ 2004b). Helium 
pumps and heaters allow thermal control of the cryogenic  components which 
are kept at 2-8~K during the critical observations.

To minimize instrumental systematic effects the horns are cooled to 
approximately the temperature of the CMB (2.7 K). The horns have a $16^\circ$ 
full-width-half-maximum beam and are pointed $30^\circ$ from the zenith to 
minimize acceptance of balloon and flight train emission. A helium cooled flare 
reduces contamination from ground emission. No windows are used. Air is kept 
from the instrument by the efflux of helium gas.

%------------------------------------------------------------------
% Table 1: Previous measurements
%------------------------------------------------------------------
\begin{table}[b]
\begin{center}
\caption{\label {measure}
Previous low frequency CMB measurements and their uncertainties 
from the literature.}
\begin{tabular}{cccl}
% \tablewidth{7.5in}
\hline
Temperature & uncertainty & frequency & source \\
K & mK & GHz & \\
\hline
2.783 & 25 & 25  & Johnson \& Wilkinson, 1987 \\ 
2.730 & 14 & 10.7& Staggs \etal, 1996b \\
2.64 & 60  & 7.5 & Levin \etal, 1992 \\
2.55 & 140 &  2  & Bersanelli \etal, 1994 \\
2.26 & 200 & 1.47& Bersanelli \etal, 1995 \\
2.66 & 320 & 1.4 & Staggs \etal, 1996a \\
3.45 & 780 & 1.28& Raghunathan \& Subrahmanyan \\
\hline
\end{tabular}
\end{center}
\end{table}

The Dicke switch chops between the reference and the horn antenna at 
100~Hz. Following the HEMT amplifier, outside the dewar, each radiometer has 
a warm amplifier followed by narrow and wide filters terminated by two 
detectors. The detectors are followed by lockin amplifiers running
synchronously with the Dicke switch. The bandwidths of the 10~GHz narrow
and wide and the 30~GHz narrow and wide radiometers are: 0.1, 1.1, 1.0, and
2.9~GHz respectively. The raw sensitivities of these 4 radiometer channels 
are: 2.3, 0.7, 4.8, and 2.8~mK$\sqrt{{\rm Hz}}$ respectively. The instrument 
details are discussed by Kogut \etal\ (2004a).
 
Other instrumentation on the ARCADE includes a magnetometer and tilt meter
to determine the pointing of the radiometers. A GPS receiver collects position
and altitude information. Thermometers, voltage and current sensors collect 
information on the condition of the instrument. A commandable rotator allows 
rotation of the entire ARCADE gondola at $\sim0.5$~RPM. A commandable tilt 
actuator allows tipping of the dewar with the radiometers to measure residual 
beam and atmospheric effects. A camera provides in flight video pictures of
the aperture plane. The camera transmitter interferes with the
radiometers so the data taken while the camera is on are not used for science.

The sky temperature estimate depends critically on the measurement of the 
calibrator temperature. The other components (horn, switch, cold reference, and
amplifier) become merely a transfer standard to compare the measurements when 
looking at the sky to the measurements when looking at the calibrator. The 
calibrator is addressed by Kogut \etal\ (2004b) in a companion paper. There 
are 7 thermometers embedded in the calibrator and its thermal buffer plate:
3 at the tips of the calibrator (T1, T2, T3), 2 at the base of the calibrator 
(T4, T5) and 2 on the control plate (B1, B2). These ruthenium oxide 
thermometers (Fixsen \etal\ 2002) are read out at approximately 1~Hz 
with .15~mK precision and 2~mK accuracy (both are poorer at temperatuers above
2.7~K). The thermometers have been
calibrated on four separate occasions over 5 years with absolute
calibrations stable to 2~mK. In addition to the NIST standard thermometer,
the lambda transition to superfluid helium at 2.18~K is clearly seen in the 
calibration data providing an absolute {\it in situ} reference.

%-----------------------------------------------------------
% Figure 2: Temperature overview
%-----------------------------------------------------------
\begin{figure}[t]
\includegraphics[width=3.5in]{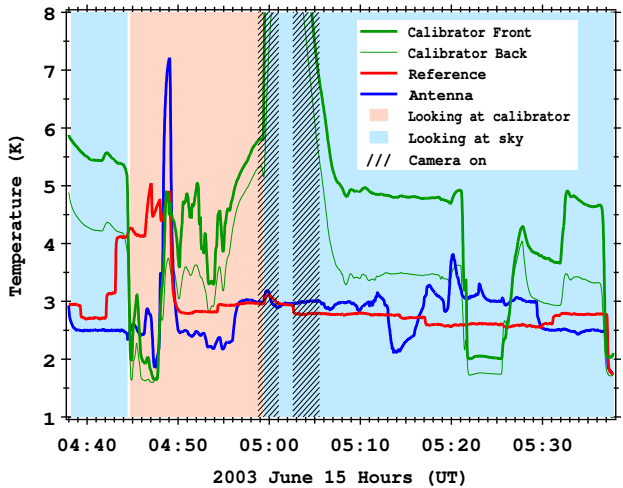}
\caption{The temperatures of critical components 
for the 10 GHz radiometer during the flight.
Solid lines indicate component temperatures.
Vertical bands indicate whether the 10 GHz antenna
viewed the sky or the external calibrator.
Cross-hatching shows times when the camera was turned on;
data from these times are not used for science analysis.
}
\label{temps}
\end{figure}

\section{THE OBSERVATIONS}
The ARCADE instrument was launched from Palestine TX on a SF3-11.82 balloon
2003 Jun 15 at 1:00~UT after an engineering launch the previous year. At
approximately 2:20~UT the instrument reached 21~km so the helium became 
superfluid. At 2:50~UT the superfluid fountain effect pumps were engaged to 
cool the upper components of the radiometer and the external calibrator. 

The ARCADE reached a float altitude of 35~km at 4:00~UT. The cover protecting
the cryogenic components was opened 
at 4:04~UT. Radiometer gains were adjusted as the external calibrator cooled 
and were set at their final values at 4:27~UT. The calibrator was moved to the 
30~GHz radiometer at 4:32~UT. After some calibration of the 30~GHz radiometer
the calibrator was moved to the 10~GHz radiometer at 4:44~UT. The next 10 
minutes provide the key 10~GHz calibration data. At 4:54~UT the calibrator was 
moved to the 30~GHz radiometer. The camera was turned on to verify this move 
and turned off at 5:00~UT.

A command was sent to move the calibrator back to the 10~GHz radiometer at
5:07~UT and the camera was turned on to verify the move. However, the calibrator
motor failed and no further moves of the calibrator were possible. At 5:11~UT
the camera was again turned off. At 5:20~UT tipping maneuvers commenced and
continued until 5:36~UT.

The helium was exhausted by 7:20 and the ARCADE continued to operate until 
8:30~UT but without the ability to move the calibrator only engineering data 
were collected after 5:36.
The most useful observations were from 4:36~UT to 5:36~UT and all of the 
following derivations will use various subsets of this data.
Some of the temperatures for this time are shown in Figure \ref{temps}.

\section{ANTENNA TEMPERATURE ESTIMATION}
Conceptually the calibration process is straightforward. The calibrator is 
placed over the aperture of a radiometer, and the four cryogenic components of 
the radiometer (horn, Dicke switch, reference, and amplifier) are each warmed 
and cooled to allow the measurement of the coupling or emission from each
component into the radiometer. The calibrator temperature is also changed to
measure the responsivity of the radiometer. The calibrator is then moved away 
so the radiometer observes the sky; the parameters measured while observing 
the calibrator are used to deduce the antenna temperature of the sky.

The most efficient use of the data uses all of the component temperature
variations to obtain the best emissivity estimations,
so a general least squares fit is used to solve for all of the emissivities 
simultaneously. By adding the assumption that the sky is a stable but unknown
temperature, the fit can take advantage of the component temperature changes 
during the sky observations as well.
The corrections for instrument and Galactic foreground are included in the 
least squares solution. 

Ideally the temperature of the cryogenic components is near that of the CMB.
A linear model can then be used to predict the radiometer output:
\begin{equation}
R=g E\cdot T
\end{equation}
where $g$ is the responsivity of the radiometer, $T$ is the matrix of 
temperatures (each row is a component, each column a time), $E$ is the vector 
of emissivities and $R$ is a vector of radiometer readings. Since neither $g$ 
nor $E$ is known a priori, they are combined; $A=gE$. A least squares fit:
\begin{equation}
A=(T\cdot W\cdot T^T)^{-1}\cdot T^T\cdot W\cdot R
\end{equation}
produces the weighted solution to the optimum parameterization $A$,
where $W$ is a weight matrix.

The list of temperatures in $T$ in the full fit includes: Five RuO thermometers
embedded in the absorber of the external calibrator, two thermometers in the 
external calibrator thermal control plate, a thermometer on the horn antenna, 
a thermometer in the reference load, a thermometer on the Dicke switch, and a 
thermometer on the cryogenic HEMT amplifier. These 11 thermometers are 
augmented by an offset and a derivative of the radiometer reading to allow a 
fit for the precise phase between the radiometer and thermometer sampling. 
Thus the full T matrix is 13$\times N$, where $N$ is  the number of observations.

Since the HEMT amplifier follows the Dicke switch it cannot affect the offset
of the radiometer, but its gain can affect the output. To correct for any 
temperature dependence in the gain, the row for the amplifier in the temperature
matrix contains $R*\delta T_{amp}$ where $\delta T_{amp}=T_{amp}-\left< 
T_{amp} \right>$ The mean of the temperature is removed to improve the condition
of the matrix which would otherwise have a row nearly identical to the 
data being fit.

Data analysis revealed a cross coupling from the 30~GHz narrow channel to
the 10~GHz wide channel which followed it in the digitization multiplexor. The 
coupling was 1.745\%, it is statistically significant and was removed before 
proceeding with the rest of the data analysis. The other data were checked for 
anomalous couplings and no others were found at the 0.003\% level, although 
couplings between the 10~GHz channels or between the 30~GHz channels would 
not appear in this analysis since they share the same front end.  The lockin 
and following electronics were 
retested after the flight but the coupling was not reproduced, suggesting
a ground loop in the original electrical configuration. The gondola was
substantially disassembled to facilitate recovery so recreating the full
original electrical configuration is not possible.

%------------------------------------------------------------------
% Table 2: Model Parameters
%------------------------------------------------------------------
\begin{table}[t]
\begin{center}
\caption{\label{fit}
Normalized Coefficients\tablenotemark{a}}
\begin{tabular}{lrrrr}
\hline
Thermometer  & 10 GHz & Radiometer & 30 GHz & Radiometer \\
Component    & Wide & Narrow & Wide & Narrow \\
\hline
Calibrator T1&  .556 &  .545 &  .720 & .854 \\
Calibrator T2&  .365 &  .371 &  .442 & .458 \\
Calibrator T3&  .059 &  .069 &  .062 & .041 \\
Calibrator T4& -.052 & -.064 & 1.306 &-.736 \\
Calibrator T5& -.033 & -.034 &-1.475 & .374 \\
Buffer B1    &  .190 &  .180 &  .405 & .217 \\
Buffer B2    & -.084 & -.067 & -.460 &-.208 \\
Horn Antenna &  .004 &  .010 & -.021 &-.001 \\
Dicke Switch &  .002 &  .007 &  .355 & .169 \\
Reference    &-1.045 &-1.041 &-1.008 &-.857 \\
HEMT Amplifier& .009 &  .007 & -.007 & .016 \\
Time shift   &  .127 &  .093 & -.599 &-.600 \\
Offset       &  .141 &  .007 & -.670 &-.941 \\
\hline
\end{tabular}
\tablenotetext{a}{Coefficients $A$ from the least squares fit (eqn 7). Coupling 
parameters have been renormalized so the sum over the 
calibrator is 1. The offset is in Kelvins, time shift is in seconds, 
HEMT amplifier is gain in K$^{-1}$ All other lines are dimensionless 
emissivities. The locations for each of the thermometers within the 
calibrator are shown by Kogut (2004b).}
\end{center}
\end{table}

The weight matrix $W$ must be chosen with some care. The weight matrix is
assumed to be diagonal. Some data are excised by making the weight zero.
Data are excised while the lockin amplifier was driven out of range
(8\% of 10~GHz and 25\% of 30~GHz data). Data were excised while the 
camera transmitter was on since the camera transmitter was shown to interfere
with the warm amplifiers before launch (10\% of the data). An additional
4\% of the 10~GHz and 15\% of the 30~GHz data were excised as outliers.
These points are all at sharp transitions where the temperature change
was rapid and the second derivative was also high. The remaining
data were weighted by $W=1/[T_x'^2+T_r'^2+.008T_x+.002T_r]$, where $T_x$
and $T_r$ are the temperatures of the external calibrator and the reference
respectively. The higher temperature data are deweighted because the 
thermometer readout noise and absolute accuracy both degrade at higher 
temperature (Kogut \etal\ 2004b, Fixsen \etal\ 2002).

The time derivatives of the temperatures are included in estimating the weight 
because the thermometer readout and the lockin readout used separate clocks, 
leading to phase jitter of approximately one second in the final data stream. 
In addition the thermal time constants are not negligible and the thermometers 
may lead or lag the emissive parts of the components. The time constants of the 
components vary from a fraction of a second for the amplifier, which is mostly 
copper, to about 5 seconds for the external calibrator which has most of its 
thermal mass in the form of Eccosorb. The time constants are also functions 
of temperature, generally becoming faster at lower temperature.

After the solution $A$ is found the responsivity can be taken out by 
renormalizing so that the net response to the external calibrator is unity.
The emissivities $E$ for the four radiometers are shown in Table~\ref{fit}. The 
``emissivity" for the internal reference should be near -1 since the reference 
is observed during the negative phase of the lockin. The calibrator temperatures
T4 and T5 are highly correlated as the thermal resistance and the heat flow 
between them is small. The value of the sum of their emissivities is thus much 
more stable than the value of the difference. Part of this results from mathematical 
instability in the inversion of the matrix and part from using the data 
to extrapolate along a gradient rather than interpolate. The same is true for 
the buffer temperatures.

%------------------------------------------------------------------
% Table 3: Component temperatures
%------------------------------------------------------------------
\begin{table}[b]
\begin{center}
\caption{\label{meantemps}
Mean Component Temperatures\tablenotemark{a}}
\begin{tabular}{l c c c c}
\hline
Component   & \multicolumn{2}{c}{10 GHz Radiometer} & 
	      \multicolumn{2}{c}{30 GHz Radiometer} \\
Temperature & mean & variation & mean & variation \\
\hline
Calibrator   & 3.94 K & .88 K & 4.16 K & 1.05 K \\
Horn Antenna & 2.79 K & .32 K & 6.93 K & 2.67 K \\
Dicke Switch & 2.62 K & .31 K & 2.89 K & 0.61 K \\
Reference    & 2.75 K & .24 K & 2.97 K & 1.34 K \\
HEMT Amplifier & 2.73 K & .17 K & 2.84 K & 0.59 K \\
\hline
\end{tabular}
\tablenotetext{a}{Temperatures and variations are mean weighted temperature
and RMS weighted variations used in the model.}
\end{center}
\end{table}

Table \ref{fit} shows that most of the components have a small coupling to the 
radiometer output. The critical components are the reference, and two of the 
external calibration thermometers near the tips of the external calibrators 
eccosorb load and near the center of the beam (T1 and T2), where most of the 
emission is expected to originate. The horn emissivity is 0.8\% or about what 
is expected from an aluminum antenna of this type at this frequency. The 
reference load emissivity is near its ideal of -1 for the 10~GHz radiometers, 
giving confidence in the fit. 

As can be seen from Table~\ref{meantemps}, the mean temperatures of 
the major components of the 10~GHz radiometer are near the CMB temperature.
This minimizes the effects of reflections, unmodeled emission and
responsivity variations.
The cold reference has as much impact on the radiometer signal
as the sky or external calibrator. But the sky temperature estimation
does not depend on the {\it absolute} accuracy of the reference thermometer.
Nevertheless the reference thermometer (like all of the thermometers) is read 
out to a precision of .15~mK and has been calibrated to 2~mK, at the lambda
point, against an absolute NIST standard.

%-----------------------------------------------------------
% Figure 3: Temperature residuals
%-----------------------------------------------------------
\begin{figure}
\includegraphics[angle=90,width=3.25in]{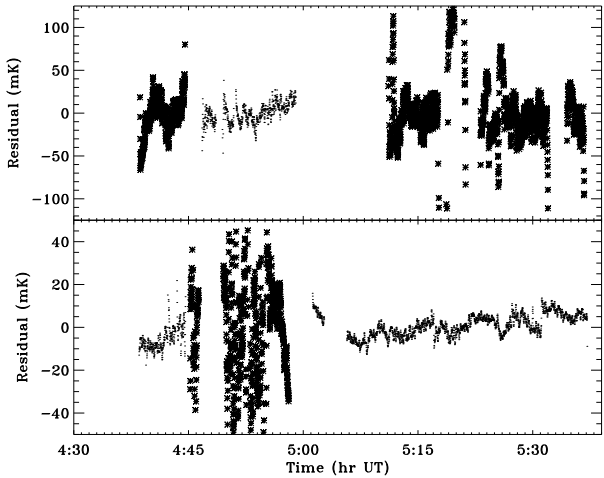}
\caption{The residuals of the fit to the calibration data for the 30 GHz
channel (top) and the 10 GHz channel (bottom). The dots are sky observations 
and the stars are external calibrator observations.}
\label{residuals}
\end{figure}

Figure~\ref{residuals} shows the residuals in the data after the model has 
been removed.
Comparing Fig.~\ref{temps} with Fig.~\ref{residuals} it can be seen that while 
the radiometer component temperatures vary by several kelvins, the residuals of 
the fit vary by only tens of millikelvins. The sky data is even better with 
only 5~mK RMS residuals. While this may be surprising for a linear model which 
has only 13 parameters and no explicit information about the radiometer, it 
demonstrates that the system is close to linear and all of the major components 
are measured. The higher residuals during the calibration are mainly due to 
the high rates of temperature change in the calibrator.

The 30~GHz data has higher residuals than the 10~GHz data 
(Fig \ref{residuals}b). There are four 
contributing reasons. First the 30~GHz radiometers have higher intrinsic noise which 
is amplified by the fitting process. Second, most of the data is calibration 
data which has additional noise from the calibrator and does not constrain some 
of the parameters of the fit as well. Third, the temperatures of the calibrator
and the antenna are higher than for the 10~GHz data providing a poorer 
match to the CMB. Fourth, the antenna observes a smaller section of the 
calibrator so gradient and changes in time are not smoothed as well as
in the 10~GHz radiometer.

To estimate the sky temperature, the temperature of all of the calibrator 
thermometers in the fit is replaced with a test sky temperature plus foreground 
models for times that the radiometer observes the sky. The test sky temperature 
is then varied to find a minimum in
\begin{equation}
\chi^2=(A \cdot T-R)\cdot W\cdot (A \cdot T-R)
\end{equation}
which is interpreted as the 
best fit sky temperature. The variation in the $\chi^2$ is then used to 
determine the statistical uncertainty of the sky temperature. The $\chi^2$ is 
renormalized at the best fit solution, so this procedure includes some, but 
not all, of the systematic effects.

\begin{figure}[t]
\includegraphics[angle=90,width=3.25in]{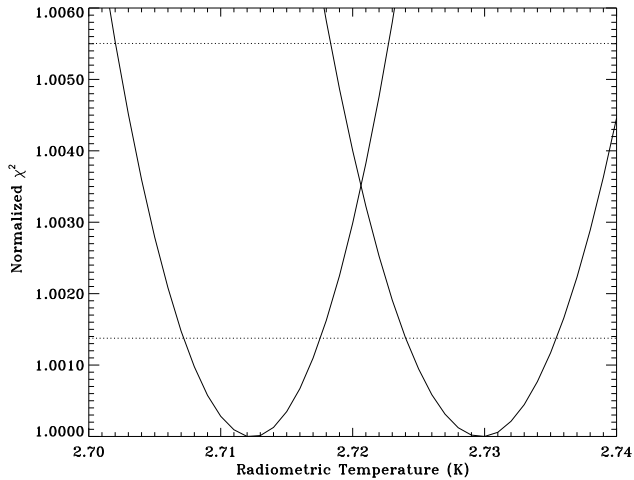}
\caption{The $\chi^2$ as a function of the test sky temperature for 
the 10 GHz channels. The parabola on the left is for the wide channel and the 
parabola on the right is for the narrow channel.  Both have been renormalized
to a $\chi^2$/DOF of 1 at the minimum. 1 and 2~$\sigma$ cutoffs are shown
as dotted lines.}
\label{Sky}
\end{figure}

In the long wavelength limit the antenna temperature and the thermodynamic
temperature are identical. For some of the lower temperatures in this 
experiment, this approximation
is marginal, so the thermodynamic temperature is translated to an antenna 
temperature for each of the measurements and the entire fit is done with
antenna temperatures. This makes a 1~mK correction to the 10~GHz result
and lowers the total $\chi^2$ by 2.6 and 1.7, for the two channels.
The effect is larger (8~mK) for the 30~GHz result with $\chi^2$ changes of
18 and 24 for the two channels.
The results of this fit is shown in Fig. \ref{Sky}.

The final thermodynamic temperatures for the four channels are:
10~GHz narrow: $2.730\pm.005$~K, 10~GHz wide: $2.712\pm.005$~K,
30~GHz narrow: $2.680\pm.016$~K, and 30~GHz wide: $2.697\pm.008$~K,
where the $1\sigma$ uncertainties are calculated from the change in $\chi^2$.

\section{FOREGROUND ESTIMATION}
One of the principal advantages of a balloon flight is that it puts the 
instrument above about 99.5\% of the atmosphere and a larger fraction
of the water vapor.  The residual atmosphere is less than 
1~mK in the 10~GHz channel (Staggs, 1996a). This is too small to 
be seen in our tipping scans; no correction is made but 1~mK uncertainty 
is included in the final uncertainty estimate.

\subsection{Estimation of the Instrument foreground}
Most of the instrument was in the far sidelobes of the horns so its thermal
emission to the radiometer is negligible.  However the flight train,
consisting of the parachute, ladder, FAA transmitter and balloon is directly 
above the radiometer only 30$^\circ$ from the center of the beam. Its emission 
could not be ignored. Since this system is complicated and moves with the 
balloon rather than the gondola, a reflector attached to the gondola was 
constructed of aluminized foam board to hide these components and instead 
reflect the sky into the radiometers.

The 4$\pi$ ~steradian antenna pattern was carefully measured in the Goddard 
test range in the flight configuration in the flight Dewar including the 
external calibrator. This measured pattern was convolved with the positions
and emissivity estimates of the flight train and balloon to estimate the
radiation from the balloon and flight train. The total expected emission does not 
change much because of the reflector, but it is much easier to compute and more
stable. The details of this calculation are provided by Kogut \etal\ (2004a).

Tip scans in flight provide a direct emission measurement from the instrument.
After the external calibrator 
and sky were observed, the radiometer was tipped up to $6^\circ$ leading to 
predictable changes in the antenna temperature as the angle to the reflector 
changed. The predicted model was fit to the tip scan data with a single 
overall scale factor, with the best fit $1.5\pm 0.3$ of the predicted signal
for the 10~GHz channels. This is within the uncertainty of the emissivity
assumed by the model.

Because of the failure in the external calibrator moving mechanism the 
measurement could not be repeated for the 30~GHz channel. But the rough 
agreement between the model and the measurement gives confidence to the 
estimate of 10~mK radiation from the reflector. We correct the data for 
1.5$\times 10$ mK$= 15$~mK and assign a $0.3\times 10$ mK$=3$~mK uncertainty 
to the correction.

%-----------------------------------------------------------
% Figure 4: chi-squared of sky temperature
%-----------------------------------------------------------

\subsection{Estimation of the Galactic foreground}
The antennas viewed the sky at Galactic latitude $13\arcdeg < b < 83\arcdeg$
with a majority of observations at $b > 35\arcdeg$. The Galactic foreground 
is estimated using models of the synchrotron, free-free, and dust emission 
derived from the Wilkinson Microwave Anisotropy Probe (WMAP) and other data 
(Bennett \etal\ 2003). The WMAP maximum entropy foreground models for each 
component are scaled to the center frequency of each ARCADE band using the
spectral index derived from the WMAP 23 and 33 GHz data. Then the components 
are combined and convolved with the symmetrized ARCADE beam pattern to 
produce a smoothed map at each ARCADE band. The data are corrected to the 
mean sky temperature by including the CMB dipole as an additional 
``foreground''. Higher order anisotropies in the CMB are insignificant at 
this level and have been ignored.

Spatial structure in the foreground model is dominated by the Galactic plane
and the CMB dipole. This spatial variation is used as a rough check of the 
model and the instrument pointing. A magnetometer and redundant clinometers 
mounted on the Dewar allow reconstruction of the pointing within 3\arcdeg.
The calibrated data are compared to the foreground model as the gondola 
rotation sweeps the beams across the sky. The calibrated data at 10 GHz
is fit to a sky model that includes a scaled version of the predicted 
foreground plus a pointing offset. The best-fit amplitude is $1.07 \pm 0.20$ 
times the predicted foreground with pointing offset less than 5\arcdeg.
                                                
Absolutely calibrated single frequency maps are unable to distinguish between 
the CMB and the spatially homogenous part of foreground emission. The 
homogenous part of each Galactic foreground is estimated using template maps 
dominated by each component. Synchrotron emission dominates the 408~MHz 
survey (Haslam \etal\ 1981). A fit to $csc(|b|)$ for $|b| > 40\arcdeg$ yields 
a zero level $20 \pm 3$ K at the Galactic poles at 408~MHz. The spectral index 
$\beta$ between 408~MHz and the ARCADE frequency bands varies across the 
sky and is not precisely known; estimates typically range from -2.7 to -3.2
(Platania \etal\ 1998, Bennett \etal\ 2003, Finkbeiner 2003b).
Uncertainty in the synchrotron spectral index dominates the uncertainty 
in the foreground zero level for ARCADE. Simply scaling the synchrotron 
zero level as $T ~ (\nu / {\rm 408~MHz})^\beta$ with $-3.2 < \beta < -2.7$
yields values between 0.7 and 3.5~mK at 10 GHz. Bennett \etal\ present a 
synchrotron model which explicitly takes into account the spatial variation
of the spectral index. Extrapolating this model using the WMAP data at 33 
and 23 GHz yields a zero level of 1.4 mK at 10.1 GHz. Since WMAP provides 
high quality data at frequencies at or near the ARCADE data, we adopt the 
WMAP model for the zero level and assign uncertainty 2 mK at 10 GHz
to account for the uncertainty in the spectral index.

Microwave free-free emission from ionized gas can be traced using H$\alpha$ 
emission from the same gas (Finkbeiner 2003a). H$\alpha$ emission at 
$|b| > 75 \deg$ has intensity 0.5 Rayleighs, corresponding to 0.04 mK at 10 GHz
(Bennett \etal\ 2003). Thermal dust emission is similarly faint, with 10 GHz 
antenna temperature below 1 $\mu$K at $|b| > 75 \deg$
(Finkbeiner, Davis, \& Schlegel 1999). All Galactic foregrounds are 
negligible (less than 0.5~mK) at 30 GHz.

\section{UNCERTAINTY ESTIMATION}
The overall uncertainty in the radiometric temperature is a combination of the
uncertainties of the parts of the model that go into the radiometric temperature
estimate. Each of the uncertainties listed in Table~\ref{uncertain} will be 
discussed in turn.

\subsection{Statistical uncertainty}
The statistical uncertainty is derived directly from the data. 
After the residuals are computed the $\chi^2$ is renormalized so that the 
$\chi^2$/DOF is one. This is larger than the radiometer noise because the 
residuals are still contaminated by some residual systematic effects.
The difference between the 10~GHz wide and 10~GHz narrow results is somewhat 
larger than one would expect, but the probability of getting a difference this 
large is 10\%. We note this is not so improbable, so the two results are 
combined into a single average. The 30~GHz data have larger uncertainties 
because of the larger noise and shorter sky observation. They too are averaged. 

\subsection{Absolute thermometer calibration uncertainty}
The sky temperature can not be determined to better accuracy than the absolute 
calibration of the thermometersin the external target. The thermometer 
calibration was tested several times before the flight and retested after the 
flight. Each test reproduces the calibration to about 2~mK accuracy verified 
against a NISTstandard and cross-checked at the $\lambda$ transition. The 
accuracy is degraded at higher temperatures, but the calibration has been shown 
to be stable over long periods. The uncertainty is larger for the 30~GHz 
because the calibrator has a higher mean temperature in the 30~GHz calibration.

%-----------------------------------------------------------
% Figure 5: Temperature residuals vs calibrator gradients
%-----------------------------------------------------------
\begin{figure}[b]
\includegraphics[angle=90,width=3.25in]{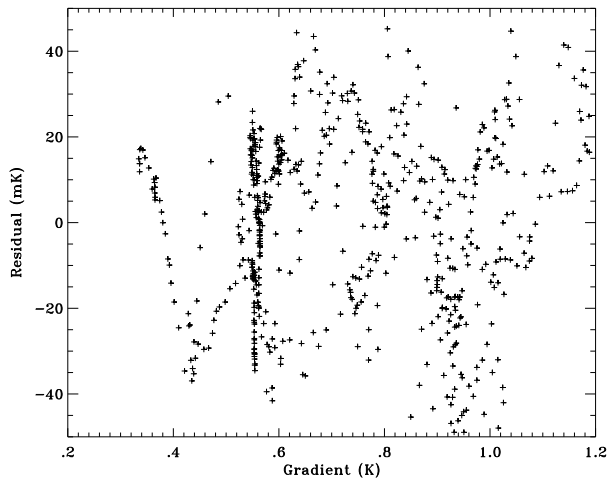}
\caption{Scatter plot of the residuals vs the front to back thermal gradient
in the calibrator. The residuals are uncorrelated with the temperature
gradients.}
\label{grad}
\end{figure}

\subsection{Temperature gradient uncertainty}
The uncertainty in the sky temperature is dominated by thermal gradients in 
the external calibrator. If the calibrator were isothermal, its only 
contribution to the sky temperature uncertainty would be the absolute 
calibration uncertainty of the embedded thermometers. Spatial gradients are 
observed within the Eccosorb absorber. The largest gradient averages 720 mK 
front-to-back, with the absorber tips warmer than the back. Transverse 
gradients are smaller, with a mean gradient of 213 mK between thermometers 
T1 and T2 and 65~mK between T2 and T3. These gradients are not stable in time,
but vary with scatter comparable to the mean amplitude.

The radiometric temperature of the external calibrator depends on the integral 
of the temperature distribution within the absorber, weighted by the electric 
field at the antenna aperture. This integral is approximated as a linear 
combination of the 5 thermometers imbedded in the absorber. The time variation 
in the temperatures and radiometer output is used to derive a single 
time-averaged weight for each thermometer. The procedure is insensitive to 
thermal gradients in directions not sampled by the thermometers,
or on spatial scales smaller than the spacing between thermometers.
The flight data are used to estimate these residual effects.

Differences between the calibrator radiometric temperature and the linearized 
model will appear as residuals in the calibrated data.
Figure 3a shows the time-ordered residuals as the 10 GHz radiometer
observes both the calibrator and the sky.
The residuals have standard deviation of 20 mK during calibration,
compared to 5~mK during observations of the sky (the weighted average is 6~mK).
Figure 5 shows the calibration residuals sorted with respect 
to the main front-to-back gradient in the calibrator.
There is no correlation between the residuals 
and the measured temperature gradients.
The uncertainty in the measured radiometric temperature
of the calibrator is thus
\begin{equation}
\delta T = \sigma_{\rm cal}/\sqrt N_{\rm eff}
\end{equation}
where
$\sigma_{\rm cal}$ is the standard deviation of the calibration residuals
and $N_{\rm eff}$ is the number of independent observations.
The calibration residuals are highly correlated in time,
so $N_{\rm eff}$ is smaller than the number of data points during calibration.
We estimate the uncertainty by noting that the 10 GHz calibration data
shows 16 zero crossings, so that
$\delta T ~\approx$ 20 mK / $\sqrt{16}$ = 5 mK at 10 GHz
and $\delta T ~\approx$ 30 mK / $\sqrt{36}$ = 5 mK at 30 GHz.

The linear model assumes a sufficient density of thermometers to adequately 
sample gradients within the absorber. This assumption is tested by dropping 
each thermometer in turn from the fit and repeating the analysis.
Additional fits derive the sky temperature after dropping two thermometers.
While the $\chi^2$ increases for these fits (sometimes dramatically),
the sky temperature remains in a narrow range provided at least one of (T1,T2) 
and at least one of (T4,T5) are included in the fit. These are the thermometers 
needed to sense the main front-to-back temperature gradient in the calibrator.
Any additional thermometer is sufficient to sample the remaining (small) 
transverse gradients, demonstrating that the 5 imbedded thermometers provide 
adequate spatial sampling of any thermal gradients within the calibrator.
The scatter in the set of solutions without one or two thermometers
serves as a conservative estimate for the uncertainty from finite sampling of 
the temperature within the calibrator. The standard deviation of the sky 
temperature for these solutions is 8 mK for the 10 GHz radiometer
and 30 mK for the 30 GHz radiometer (which has a much smaller ``footprint'' of 
the antenna aperture on the calibrator). This is comparable to the uncertainty 
derived from the time-ordered data using a radically different approach.
We conservatively adopt the larger value as our estimate for the uncertainty 
resulting from temperature gradients within the calibrator.
\subsection{Instrument emission uncertainty}
The emission from the reflector and the flight train was modeled and measured
and the two agree within the measurement uncertainties. The careful measurement
of the beam from the mouth of the antenna allows
a very complete model. The tipping tests demonstrate that model is basically
correct. The major uncertainty in the model is the emissivity of the 
aluminum foil on the foam. By calibrating against the measurement this 
uncertainty is reduced.

%------------------------------------------------------------------
% Table 4: Uncertainties
%------------------------------------------------------------------
\begin{table*}[t]
\begin{center}
\caption{\label {uncertain}
Uncertainty Summary}
\begin{tabular}{l c c l}
\hline
 	& 10 GHz 	& 30 GHz 	& \\
Source 	& Radiometer 	& Radiometer 	& Notes \\
\hline
Statistical Uncertainty & 4 mK & 7 mK & from $\chi^2$ \\ 
Thermometer Calibration & 3 mK & 4 mK & from lab tests \\
Calibrator Gradients    & 8 mK & 30 mK & fits with thermometers omitted\\
Instrument Emission     & 3 mK & 5 mK & 30\% of model \\
Atmosphere Emission     & $<1$ mK & $<1$ mK & signal is smaller than this \\
Galactic Emission       & 2 mK & $<1$ mK & Synchrotron zero point uncertainty \\
\hline
Total Uncertainty       & 10 mK & 32 mK & \\
\hline
\multicolumn{4}{c}{Uncertainty estimates are discussed in \S 6.
Uncertainties are added in quadrature.} \\
\end{tabular}
\end{center}
\end{table*}

\subsection{Atmosphere and Galactic emission uncertainty}
The atmosphere has minimal contribution at 35~km, and Galactic emission
at the radiometer frequencies is small out of the Galactic plane. Even large
fractional uncertainties in these estimations are unimportant to the final 
measurement.

\section{DISCUSSION}
Since the sky data are compared directly to the external calibrator, systematic
effects from the instrument are all eliminated to first order. Instead the
burden falls on the external calibrator. The key issues are the emissivity
(or blackness) of the calibrator, the accuracy of the thermometers and the
relationship between the temperature of the thermometers and the temperature
of the emitting surfaces.

The ARCADE external calibrator has been measured to be $>99.97$\% emissive at
10~GHz using the flight horn antenna. The emissivity measurement was
done at 295~K but the change from 295~K to 3~K should be small because the 
resistance changes less than 50\% from 300~K to 1~K (Hemmati \etal 1985).
But the reflected radiation is almost entirely radiation from the horn which
was at $\sim2.8$~K. The residual uncertainty from this effect is 0.1~mK.

The remaining issue is the relationship between the temperature of the 
thermometers and the temperature of the emitting surface of the external 
calibrator. If there were no gradients the issue would vanish, but typically
the gradient is 700~mK during the time of the observations. Some of these
gradients are measured showing that the main gradient is from the front to the 
back of the external calibrator. The mean temperature of the emitting surface 
is modeled by a linear combination of the seven thermometers on the external
calibrator. The changing temperature gradients themselves determine the 
fit to the combination of thermometers that best models the data. As long as
the variations reflect the actual mean temperatures this is a good model.
Tests fitting the sky temperature after dropping individual thermometers
demonstrate that the calibrator has enough thermometers to adequately sense
the thermal gradients.

The engineering tests done during this flight provide a base for the design
and operation of the next ARCADE mission which will have 6 frequencies extending
from 3~GHz to 90~GHz.

The ARCADE instrument has measured the radiometric temperature of the CMB to 
be ~2.721~K \err\ at 10 GHz, and 2.994~K$\pm 32$~mK at 30~GHz.  

\acknowledgements
This work was supported by the Office of Space Sciences at NASA Headquarters.
The research described in this paper was carried out in part at
the Jet Propulsion Laboratory, California Institute of Technology,
under contract with the National Aeronautics and Space Administration.

\end{document}